\documentclass[journal=jctcce,manuscript=article,layout=onecolumn]{achemso}

\usepackage{amsmath}
\usepackage{amssymb}
\usepackage[version=4]{mhchem}
\usepackage{longtable}
\usepackage{array}

\usepackage[T1]{fontenc}       % Use modern font encodings

\newcommand{\exciting}[1]{\texttt{exciting}#1}
\newcolumntype{R}[1]{>{\raggedleft\arraybackslash}p{#1}}

\author{Ronaldo Rodrigues Pela}
\affiliation{Departamento de F\'{i}sica, Instituto Tecnol\'{o}gico de Aeron\'{a}utica (ITA), 12228-900 S\~{a}o Jos\'{e} dos Campos/SP, Brazil}
\alsoaffiliation{Physics Department and IRIS Adlershof, Humboldt-Universit\"{a}t zu Berlin, Zum Gro\ss{}en Windkanal 6, D-12489 Berlin}
\alsoaffiliation{European Theoretical Spectroscopy Facility (ETSF)}
\email{ronaldorpela@gmail.com}
\author{Andris Gulans}
\affiliation{Physics Department and IRIS Adlershof, Humboldt-Universit\"{a}t zu Berlin, Zum Gro\ss{}en Windkanal 6, D-12489 Berlin}
\alsoaffiliation{European Theoretical Spectroscopy Facility (ETSF)}
\author{Claudia Draxl}
\affiliation{Physics Department and IRIS Adlershof, Humboldt-Universit\"{a}t zu Berlin, Zum Gro\ss{}en Windkanal 6, D-12489 Berlin}
\alsoaffiliation{European Theoretical Spectroscopy Facility (ETSF)}

\title{The LDA-1/2 method applied to atoms and molecules}


\begin{document}

\begin{abstract}
The LDA-1/2 method has proven to be a viable approach for calculating band gaps of semiconductors. 
To address its accuracy for finite systems, we apply LDA-1/2 to atoms and the molecules of the $GW100$ test set. The obtained energies of the highest-occupied molecular orbitals are validated against CCSD(T) data and the $G_0W_0$ approach of many-body perturbation theory. The accuracy of LDA-1/2 and $G_0W_0$ is found to be the same, where the latter is computationally much more involved. To get insight into the benefits and limitations of the LDA-1/2 method, we analyze the impact of each assumption made in deriving the methodology.
\end{abstract}

\section{Introduction}
\par A long-standing issue of approximate functionals in density-functional theory (DFT), especially the local-density approximation (LDA), is that they do not obey Koopmans' theorem. This implies that even the Kohn-Sham (KS)eigenvalue corresponding to the highest occupied molecular level (HOMO) in theory lacks a physical meaning.\cite{Chong2002,Beschtedt2015} 
%{\red should be Bechstedt in the reference above}
This problem can be resolved, e. g., by applying a self-energy correction as obtained from the $GW$ approach of many-body perturbation theory, to obtain the corresponding quasi-particle energy. It is highly desirable though to stay within the KS framework due to the lower computational costs of semi-local density functionals with its more favorable scaling with respect to the system size (third vs. fourth power of the number of atoms in the unit cell).\cite{McKechnie2015} 

While there are numerous other approaches for correcting KS eigenvalues in order to obtain improved ionization energies (IEs) \cite{Dabo2010,Ferreti2014,Zhang2015,Zheng2011,Verma2014,Zheng2013,Teale2008}, in this work, we focus on the LDA-1/2 method.\cite{Ferreira2008,Ferreira2011,Ferreira2013} It has been shown to give good results for solids, alloys, interfaces, 2D materials, and impurities.\cite{Ferreira2013,Pela2011,Pela2012,Filho2013,Santos2012,Matusalem2013,Guilhon2015} In a recent investigation,\cite{Pela2016} LDA-1/2 also proved to be a good starting point for $G_0W_0$ calculations of solids, suggesting that the method provides a good estimate to quasi-particle energies. For finite systems, its accuracy has not been assessed yet. It is therefore important to verify to which extent the LDA-1/2 method can be applied to  accurately describe the single-particle spectra in atoms and molecules. 

In this work, we perform benchmark calculations for the IEs of atoms as well for the molecules from the $GW100$ test set\cite{vanSetten2015}. For comparison, we also use the local-density approximation (LDA), the hybrid functional PBE0, and the Hartree-Fock (HF) method. To establish the accuracy of each approach, we compare our results to IEs 
obtained %with $\Delta$SCF calculations, 
in Ref. \citenum{Krause2015} with coupled-cluster calculations that include singles, doubles, and perturbative triples CCSD(T), which is considered the \emph{gold} standard among the quantum chemistry methods.\cite{Helgaker2008,Pittner} The comparison with CCSD(T) rather than with experiments allows us to safely ignore effects of temperature, nuclear vibrations, and interaction with the environment, which affect experimental values.\cite{Gallandi2015,Caruso2016} 
Further, we select 6 molecules from the $GW100$ set and determine the impact of each approximation assumed in the LDA-1/2 method. Finally, for a representative subset of 5 molecules, we evaluate the accuracy of each method in describing not only the HOMO, but also lower-lying KS levels. 

\section{The LDA-1/2 method}\label{sec-lda05}
\par The LDA-1/2 method resembles Slater's transition state technique, in which a $\Delta$SCF calculation of the IE is replaced by a single calculation of the HOMO with half-occupation. Mathematically, this reads
\begin{equation}
E(N-1)-E(N) = 
-\varepsilon_{\alpha}(N-1/2),
\end{equation}
where $E(N)$ is the total energy of a $N$-electron system, and $\varepsilon_\alpha(N-1/2)$ is the HOMO eigenvalue with half-ionization. It is possible to obtain $\varepsilon_\alpha(N-1/2)$ without explicitly removing half an electron from the molecule. This is achieved by evaluating the following inner product:\cite{Ferreira2008,Ferreira2011}
\begin{equation}
\varepsilon_{\alpha} (N-1/2) = \left\langle \phi_{\alpha}
\bigg\lvert -\frac{\nabla^2}{2}+ v_{ext}+v_{H}+v_{XC}-V_S\bigg\rvert \phi_{\alpha}
\right\rangle.\label{eq-inner}
\end{equation}
Here, we denote the HOMO wavefunction of a system with $N$ electrons as $\phi_\alpha$, while the external ($v_{ext}$), Hartree ($v_{H}$) and exchange-correlation ($v_{XC}$) potentials are the contributions to the KS potential. The term $V_S$, called self-energy potential due to its similarity to its electrostatic counterpart, carries the information about the half-ionization.\cite{Pela2016} It can be expressed as the difference between the KS potentials of the $N$ and $N-1/2$ electron systems, respectively.\cite{Ferreira2008,Ferreira2011} Conceptually, $V_S$ is the potential needed to create half a hole in the HOMO.\cite{Ferreira2008,Ferreira2011}

A good approximation to $V_S$ is to consider it as a sum of the self-energy potentials of its atoms,\cite{Ferreira2011}
\begin{equation}\label{eq-linearcombinationVS}
V_S(\mathsf{compound}) \cong 
\sum_i V_{S}(f_i,\mathsf{atom}_i),
\end{equation}
where $f_i$ is the fractional charge removed from the $i$-th atom. Each self-energy potential $V_{S}(f_i,\mathsf{atom}_i)$ is then obtained in a separate calculation, as a difference between the KS potential of the atom and its corresponding ion with charge $f_i$. $f_i$ should reflect how much each atom contributes to the HOMO of the actual system.
% (it is taken as proportional to the density of states projected on each atom). 
% CD I would not address this here. To be clear, one should say it in some more detail.
To create the half hole, the total degree of ionization must satisfy $\sum_i f_i=1/2$. Equation (\ref{eq-linearcombinationVS}) enables us to easily obtain the self-energy potential of a system, without half-ionizing it, as just the constituent atoms need to be ionized separately.

The last approximation in LDA-1/2 is to self-consistently solve the KS equations of the $N$-electron system with a modified XC potential according to 
\begin{equation}\label{eq-modifiedKSeq}
\left(-\frac{\nabla^2}{2}+
v_{ext}+v_{H}+v_{XC}'
\right)
\phi_j = \varepsilon_j \phi_j,
\end{equation}
where $v'_{XC}=v_{XC}-V_S$. This expression facilitates the implementation, since one just needs to modify the XC potential to employ the LDA-1/2 method in practice. Neglecting the changes in the KS wavefunctions of Eq. (\ref{eq-modifiedKSeq}) due to the inclusion of $V_S$, the HOMO eigenvalue from Eq. (\ref{eq-modifiedKSeq}) is equal to the inner product of Eq. (\ref{eq-inner}), and hence is equal to an eigenvalue with half-occupation.

For the sake of clarity, we enumerate all the assumptions of the LDA-1/2 method: 
\begin{enumerate}
	\item LDA is accurate enough to calculate IEs within a $\Delta$SCF procedure;
	\item A half-occupied eigenvalue is a good approximation to the IE calculated this way;
	\item The half-occupied eigenvalue can be obtained by means of an inner product given by Eq. (\ref{eq-inner});
	\item The self-energy potential can be expanded in terms of atomic self-energy potentials;
	\item Changes in the KS wavefunction are neglected, giving the HOMO eigenvalue according to Eq. (\ref{eq-inner}).
\end{enumerate}
Further below, we shall examine these points one by one to assess their effect on the accuracy of the method.

\section{Computational details}
LDA and LDA-1/2 calculations are performed with the full-potential linearized augmented planewaves (LAPW) code \exciting.\cite{Gulans2014,Pela2017,Gulans2018} Since periodic boundary conditions are used, we place the atoms and molecules in a box with a sufficient amount of vacuum to isolate the replicas. Based on our convergence tests, we can ensure a precision of the HOMO energies of about 10-25 meV.

In order to carry out PBE0 and HF calculations, we resort to the NWChem\cite{Valiev2010} code and rely on the Gaussian orbitals available in Def2-QZVPP basis.\cite{Weigend2005}
%The NWChem\cite{Valiev2010} code is employed to carry out PBE0 and HF calculations with the Def2-QZVPP basis.\cite{Weigend2005} 
Differences between this basis and LAPW have been carefully checked by comparing the LDA HOMO eigenvalues obtained with both codes. The mean absolute deviation is 28 meV.

For the molecules, we adopt here the experimental molecular geometries, previously employed by van Setten and coworkers \cite{vanSetten2015} in their $G_0W_0$ benchmark calculations. These geometries were also used in the CCSD(T) study in Ref. \citenum{Krause2015}.

\section{Results and discussion}

\subsection{HOMO energies of atoms and molecules} 

In Table \ref{tab-AllHOMO}, we present the calculated HOMO eigenvalues of the atoms and molecules from the $GW$100 test set. We also show CCSD(T) IEs, extracted from the literature\cite{Krause2015}. CCSD(T) data are not available for Xe.
%, the 5th system in the list, and, therefore, it is not considered when we analyze the deviations between our calculations and CCSD(T).

\begin{table*}[htbp]
\caption{HOMO energies (in eV) of atoms and molecules calculated with LDA, PBE0, HF, and LDA-1/2, and the IEs (in eV) obtained by means of $\Delta$SCF calculations with CCSD(T)\cite{Krause2015}. The first column is the index of each atom/molecule as defined by van Setten \emph{et al.}\cite{vanSetten2015}.}\label{tab-AllHOMO}
\centering
\begin{tabular}{rR{2 cm}R{2 cm}R{2 cm}R{2 cm}R{2 cm}R{2 cm}}
\hline
	&				&	CCSD(T)	&	LDA	&	PBE0	&	HF	&	LDA-1/2	\\ \hline
1	&	\ce{He}	&	24.51	&	15.52	&	18.21	&	24.98	&	24.51	\\
2	&	\ce{Ne}	&	21.32	&	13.55	&	15.98	&	23.14	&	20.54	\\
3	&	\ce{Ar}	&	15.54	&	10.40	&	11.97	&	16.08	&	15.16	\\
4	&	\ce{Kr}	&	13.94	&	9.39	&	10.77	&	14.26	&	13.61	\\
5	&	\ce{Xe}	&			&	8.38	&	9.51	&		12.41	&	12.05	\\
6	&	\ce{H2}	&	16.40	&	10.26	&	12.00	&	16.18	&	15.39	\\
7	&	\ce{Li2}	&	5.27	&	3.24	&	3.79	&	4.95	&	5.03	\\
8	&	\ce{Na2}	&	4.95	&	3.23	&	3.61	&	4.53	&	4.99	\\
9	&	\ce{Na4}	&	4.23	&	2.76	&	3.08	&	3.83	&	3.84	\\
10	&	\ce{Na6}	&	4.35	&	3.10	&	3.39	&	4.07	&	3.94	\\
11	&	\ce{K2}	&	4.06	&	2.70	&	2.94	&	3.60	&	3.94	\\
12	&	\ce{Rb2}	&	3.93	&	2.64	&	2.84	&	3.42	&	3.78	\\
13	&	\ce{N2}	&	15.57	&	10.40	&	12.16	&	16.69	&	14.65	\\
14	&	\ce{P2}	&	10.47	&	7.25	&	8.10	&	10.09	&	10.20	\\
15	&	\ce{As2}	&	9.78	&	6.69	&	7.48	&	9.29	&	9.46	\\
16	&	\ce{F2}	&	15.71	&	9.64	&	11.79	&	18.13	&	14.56	\\
17	&	\ce{Cl2}	&	11.41	&	7.42	&	8.71	&	12.07	&	10.80	\\
18	&	\ce{Br2}	&	10.54	&	6.95	&	8.12	&	11.04	&	9.99	\\
19	&	\ce{I2}	&	9.51	&	6.42	&	7.37	&	9.82	&	9.07	\\
20	&	\ce{CH4}	&	14.37	&	9.47	&	11.00	&	14.84	&	13.02	\\
21	&	\ce{C2H6}	&	13.04	&	8.14	&	9.61	&	13.25	&	11.01	\\
22	&	\ce{C3H8}	&	12.05	&	7.72	&	9.17	&	12.74	&	10.40	\\
23	&	\ce{C4H10}	&	11.57	&	7.56	&	8.97	&	12.42	&	10.01	\\
24	&	\ce{C2H4}	&	10.67	&	6.96	&	7.89	&	10.29	&	10.42	\\
25	&	\ce{C2H2}	&	11.42	&	7.38	&	8.43	&	11.19	&	11.04	\\
26	&	\ce{C4}	&	11.26	&	7.35	&	8.62	&	11.49	&	10.40	\\
27	&	\ce{C3H6}	&	10.87	&	7.20	&	8.34	&	11.36	&	10.15	\\
28	&	\ce{C6H6}	&	9.29	&	6.54	&	7.29	&	9.16	&	9.01	\\
29	&	\ce{C8H8}	&	8.35	&	5.49	&	6.25	&	8.29	&	7.58	\\
30	&	\ce{C5H6}	&	8.68	&	5.60	&	6.39	&	8.41	&	8.28	\\
31	&	\ce{C2H3F}	&	10.55	&	6.74	&	7.78	&	10.49	&	10.07	\\
32	&	\ce{C2H3Cl}	&	10.09	&	6.61	&	7.58	&	10.11	&	9.57	\\
33	&	\ce{C2H3Br}	&	9.27	&	6.00	&	6.91	&	9.19	&	8.90	\\
34	&	\ce{C2H3I}	&	9.33	&	6.21	&	7.09	&	9.37	&	8.94	\\
35	&	\ce{CF4}	&	16.30	&	10.66	&	12.67	&	18.65	&	14.15	\\
36	&	\ce{CCl4}	&	11.56	&	7.80	&	9.10	&	12.52	&	10.28	\\
37	&	\ce{CBr4}	&	10.46	&	7.10	&	8.27	&	11.24	&	9.33	\\
38	&	\ce{CI4}	&	9.27	&	6.37	&	7.31	&	9.80	&	8.25	\\
39	&	\ce{SiH4}	&	12.80	&	8.52	&	9.87	&	13.24	&	11.42	\\
40	&	\ce{GeH4}	&	12.50	&	8.37	&	9.67	&	12.89	&	11.24	\\
\end{tabular}
\end{table*}

\begin{table*}[htbp]
\begin{tabular}{rR{2 cm}R{2 cm}R{2 cm}R{2 cm}R{2 cm}R{2 cm}}
\hline
	&				&	CCSD(T)	&	LDA	&	PBE0	&	HF	&	LDA-1/2	\\ \hline
41	&	\ce{Si2H6	}	&	10.65	&	7.34	&	8.40	&	11.06	&	9.62	\\
42	&	\ce{Si5H12	}	&	9.27	&	6.66	&	7.55	&	9.81	&	7.99	\\
43	&	\ce{LiH	}	&	7.96	&	4.39	&	5.44	&	8.21	&	8.13	\\
44	&	\ce{KH	}	&	6.13	&	3.58	&	4.27	&	6.56	&	6.57	\\
45	&	\ce{BH3	}	&	13.28	&	8.49	&	9.94	&	13.57	&	11.90	\\
46	&	\ce{B2H6	}	&	12.26	&	7.85	&	9.27	&	12.85	&	10.61	\\
47	&	\ce{NH3	}	&	10.81	&	6.29	&	7.70	&	11.69	&	10.52	\\
48	&	\ce{HN3	}	&	10.68	&	7.01	&	8.04	&	10.99	&	10.59	\\
49	&	\ce{PH3	}	&	10.52	&	6.78	&	7.87	&	10.58	&	9.84	\\
50	&	\ce{AsH3	}	&	10.40	&	6.89	&	7.84	&	10.39	&	10.04	\\
51	&	\ce{SH2	}	&	10.31	&	6.41	&	7.55	&	10.48	&	10.10	\\
52	&	\ce{FH	}	&	16.03	&	9.82	&	11.79	&	17.68	&	15.71	\\
53	&	\ce{ClH	}	&	12.59	&	8.15	&	9.49	&	12.97	&	12.36	\\
54	&	\ce{LiF	}	&	11.32	&	6.36	&	7.91	&	12.93	&	11.57	\\
55	&	\ce{MgF2	}	&	13.71	&	8.55	&	10.22	&	15.41	&	12.11	\\
56	&	\ce{TiF4	}	&	15.48	&	10.70	&	12.50	&	17.96	&	13.76	\\
57	&	\ce{AlF3	}	&	15.46	&	9.98	&	11.77	&	17.26	&	13.33	\\
58	&	\ce{BF	}	&	11.09	&	6.82	&	8.02	&	11.02	&	10.51	\\
59	&	\ce{SF4	}	&	12.59	&	8.46	&	9.82	&	13.81	&	11.76	\\
60	&	\ce{BrK	}	&	8.13	&	4.90	&	5.85	&	8.56	&	8.26	\\
61	&	\ce{GaCl	}	&	9.77	&	6.77	&	7.49	&	9.55	&	9.64	\\
62	&	\ce{NaCl	}	&	9.03	&	5.44	&	6.52	&	9.63	&	9.15	\\
63	&	\ce{MgCl2	}	&	11.67	&	7.78	&	9.00	&	12.25	&	10.37	\\
64	&	\ce{AlI3	}	&	9.82	&	6.81	&	7.75	&	10.18	&	8.56	\\
65	&	\ce{BN	}	&	11.89	&	7.46	&	8.73	&	11.54	&	11.40	\\
66	&	\ce{HCN	}	&	13.87	&	9.18	&	10.41	&	13.52	&	13.29	\\
67	&	\ce{PN	}	&	11.74	&	7.85	&	9.30	&	12.06	&	11.41	\\
68	&	\ce{H2NNH2	}	&	9.72	&	5.39	&	6.80	&	10.72	&	8.78	\\
69	&	\ce{H2CO	}	&	10.84	&	6.38	&	7.88	&	12.05	&	10.16	\\
70	&	\ce{CH4O	}	&	11.04	&	6.45	&	8.01	&	12.34	&	10.24	\\
71	&	\ce{C2H6O	}	&	10.69	&	6.27	&	7.80	&	12.04	&	9.82	\\
72	&	\ce{C2H4O	}	&	10.21	&	6.09	&	7.54	&	11.58	&	9.61	\\
73	&	\ce{C4H10O	}	&	9.82	&	5.92	&	7.36	&	11.39	&	8.84	\\
74	&	\ce{CH2O2	}	&	11.42	&	7.12	&	8.59	&	12.92	&	10.95	\\
75	&	\ce{HOOH	}	&	11.59	&	6.62	&	8.31	&	13.31	&	10.73	\\
76	&	\ce{H2O	}	&	12.57	&	7.41	&	9.03	&	13.87	&	12.48	\\
77	&	\ce{CO2	}	&	13.71	&	9.32	&	10.72	&	14.83	&	13.29	\\
78	&	\ce{CS2	}	&	9.98	&	6.93	&	7.85	&	10.13	&	9.78	\\
79	&	\ce{OCS	}	&	11.17	&	7.64	&	8.70	&	11.46	&	10.84	\\
80	&	\ce{OCSe	}	&	10.79	&	7.12	&	8.10	&	10.57	&	10.26	\\
81	&	\ce{CO	}	&	14.21	&	9.11	&	11.03	&	15.38	&	13.26	\\
82	&	\ce{O3	}	&	12.55	&	8.18	&	9.84	&	13.29	&	12.09	\\
83	&	\ce{SO2	}	&	13.49	&	8.29	&	9.61	&	13.53	&	11.77	\\
84	&	\ce{BeO	}	&	9.94	&	6.30	&	7.36	&	10.57	&	10.26	\\		
85	&	\ce{MgO	}	&	7.49	&	4.95	&	6.07	&	8.74	&	8.23	\\
\end{tabular}
\end{table*}

\begin{table*}[htbp]
\begin{tabular}{rR{2 cm}R{2 cm}R{2 cm}R{2 cm}R{2 cm}R{2 cm}}
\hline
	&				&	CCSD(T)	&	LDA	&	PBE0	&	HF	&	LDA-1/2	\\ \hline
86	&	\ce{C7H8	}	&	8.90	&	6.19	&	6.95	&	8.81	&	8.40	\\
87	&	\ce{C8H10	}	&	8.85	&	6.20	&	6.95	&	8.80	&	8.36	\\
88	&	\ce{C6F6	}	&	9.93	&	6.90	&	7.90	&	10.49	&	9.17	\\
89	&	\ce{C6H5OH	}	&	8.70	&	5.83	&	6.68	&	8.77	&	8.27	\\
90	&	\ce{C6H5NH2	}	&	7.99	&	5.22	&	6.03	&	8.11	&	7.57	\\
91	&	\ce{C5H5N	}	&	9.66	&	6.04	&	7.49	&	9.46	&	8.83	\\
92	&	\ce{C5H5N5O	}	&	8.03	&	5.54	&	6.22	&	8.16	&	7.54	\\
93	&	\ce{C5H5N5	}	&	8.33	&	5.76	&	6.47	&	8.39	&	7.88	\\
94	&	\ce{C4H5N3O	}	&	9.51	&	5.95	&	6.82	&	9.35	&	8.38	\\
95	&	\ce{C5H6N2O2	}	&	9.08	&	6.24	&	7.13	&	9.61	&	8.43	\\
96	&	\ce{C4H4N2O2	}	&	10.13	&	6.49	&	7.49	&	10.04	&	8.86	\\
97	&	\ce{CH4N2O	}	&	10.05	&	6.12	&	7.53	&	11.39	&	9.53	\\
98	&	\ce{Ag2	}	&	7.49	&	5.55	&	5.71	&	6.34	&	8.06	\\
99	&	\ce{Cu2	}	&	7.57	&	4.98	&	5.73	&	6.45	&	8.05	\\
100	&	\ce{NCCu	}	&	10.85	&	7.05	&	8.18	&	11.32	&	10.38	\\\hline
\end{tabular}
\end{table*}

\begin{figure}[htb]
\includegraphics[scale=0.98]{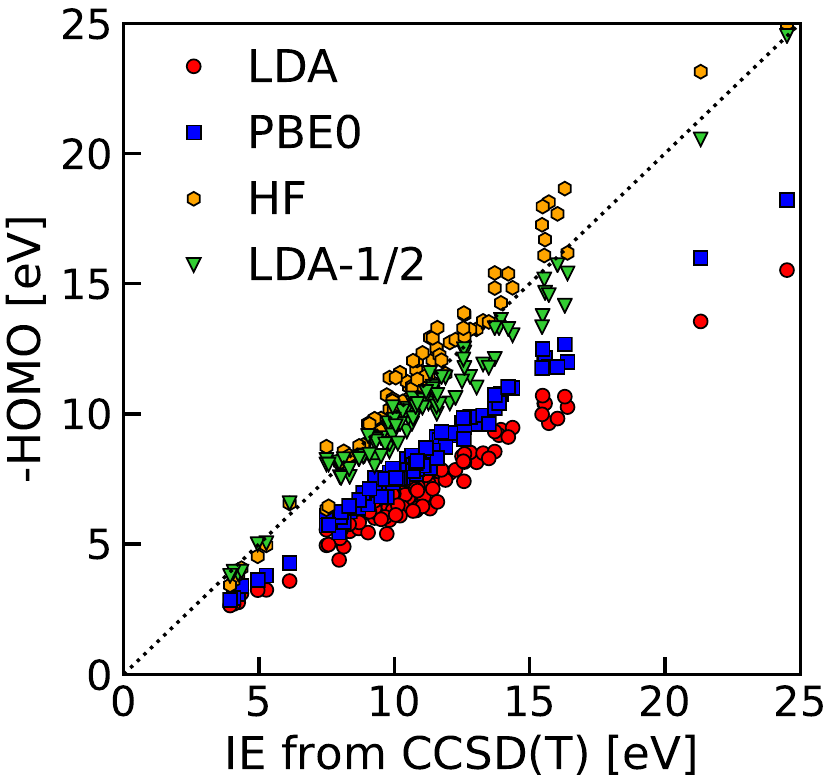}
\caption{HOMO eigenvalues as calculated with different methods in comparison with IEs from CCSD(T). The dashed line indicates perfect agreement.}\label{fig-linearfit}
\end{figure}

\begin{figure}[htb]
	\includegraphics[scale=0.98]{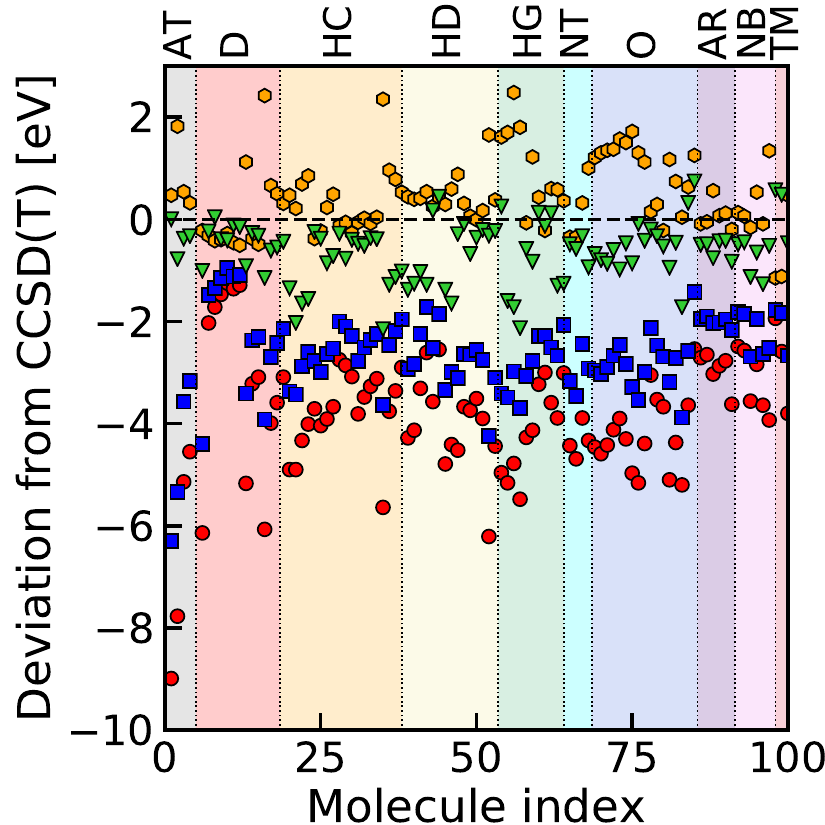}
	\caption{Deviation of each $-\varepsilon_{HOMO}$ from CCSD(T) IE. The color code is consistent with Fig. \ref{fig-linearfit}. The molecules are grouped as in Ref. \citenum{Caruso2016}.  AT: atoms, D: dimers, HC: hydrocarbons, HD: hydrides, HG: halogenides, NT: nitrides, O: oxides, AR: aromatics, NB: nucleobases, TM: transition metals.}\label{fig-error-molecules}
\end{figure}

In Fig. \ref{fig-linearfit}, we plot the HOMO eigenvalues as calculated by different functionals against the IEs from CCSD(T). The straight line corresponds to perfect agreement.
For each method, a linear fit, $y=\gamma x$, is performed to assess this accuracy quantitatively. 
The respective coefficients are given in Table \ref{tab-me}. With $\gamma=0.65$, LDA is the least accurate method. As expected, an improvement is found when PBE0 is employed, giving $\gamma = 0.76$. Finally, HF and LDA-1/2 provide even better agreement with CCSD(T). A notable difference between them is that HF tends to overestimate IEs, whereas LDA-1/2 tends to underestimate them. The level of accuracy of the two approaches is very similar though, with linear coefficients of $1.05$ and $0.94$ for HF and LDA-1/2, respectively. This means that, on average, the predictions are 5\% higher and 6\% lower, respectively, than the ideal values.

In Fig. \ref{fig-error-molecules}, the deviation of the HOMO eigenvalues with respect to CCSD(T) is plotted against the molecular index. Like in Ref. \citenum{Caruso2016}, the atoms/molecules are grouped according to their similarities. With this grouping, it is easy to identify that LDA-1/2 tends to be more accurate than HF for atoms, dimers, nitrides, oxides, and compounds with transition metals. For hydrocarbons, hydrides, halogenides, aromatic systems, and nucleobases, HF tends to be more accurate, although the accuracy of LDA-1/2 in aromatic systems is still decent (mean absolute error smaller than 0.6~eV).

\begin{figure*}[htb]
\includegraphics{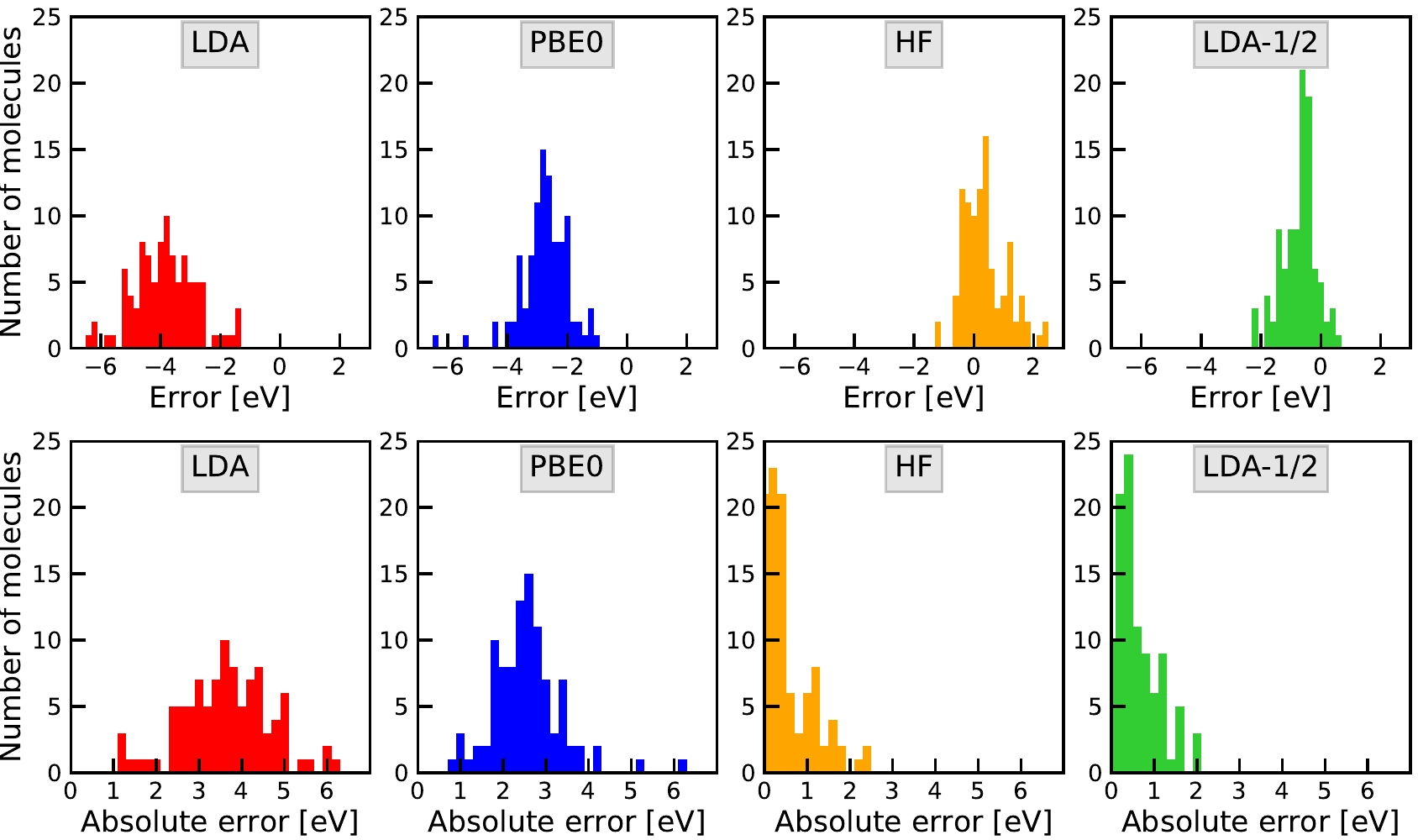}
\caption{Histograms showing the deviation of the HOMO energies obtained by different methods with respect to the CCSD(T) results of Ref. \citenum{Krause2015}. Mean errors and absolute errors are shown in the top and bottom panels, respectively.}\label{fig-HistogramAllMethods}
\end{figure*}

In Fig. \ref{fig-HistogramAllMethods}, we show histograms with the distribution of errors among LDA, PBE0, HF, and LDA-1/2. The mean errors (ME) and the mean absolute errors (MAE) are also presented in Table \ref{tab-me}. For the 100 molecules and atoms of the $GW$100 test set, the underestimation of the LDA HOMO eigenvalues compared to CCSD(T) shows up in a MAE of 3.84~eV and the same value for the ME. The HOMO levels obtained with PBE0 are also underestimated, with a MAE of 2.66~eV. As expected, the HOMO energies calculated with HF tend to overestimate IEs, while the opposite is the case for LDA-1/2. Their MAEs of 0.64~eV (HF) and 0.70~eV (LDA-1/2) are very similar. 
It is also interesting to note that the accuracy of these two methods is comparable to $G_0W_0@$PBE, which yields a MAE of 0.69~eV, as reported by Caruso and coworkers\cite{Caruso2016}. This similarity between LDA-1/2 and $G_0W_0$@PBE can be understood if we recall that LDA-1/2 is intended to reproduce the HOMO eigenvalue as the IE of a $\Delta$SCF calculation, which may reach similar accuracy to $GW$ approaches.\cite{Martin2016}

\begin{table}
\caption{MAE and ME for different approaches considered in this work. $\gamma$ is the coefficient of the linear fit through the data points in Fig. \ref{fig-error-molecules}.}\label{tab-me}
\begin{tabular}{c|ccc}
\hline
	& ME &	MAE & $\gamma$\\ \hline
LDA	&-3.84&	3.84 & 0.65\\
PBE0 &	-2.66&	2.66 & 0.76\\
HF	& 0.46&	0.64 & 1.05\\
LDA-1/2	&-0.63&	0.70 & 0.94 \\
$G_0W_0$@PBE\textsuperscript{\emph{a}} &	-0.69&	0.69 \\ \hline
\end{tabular}
\\ \textsuperscript{\emph{a}} Extracted from Ref.\citenum{Caruso2016}.
\end{table}

\begin{figure}[htb]
\includegraphics{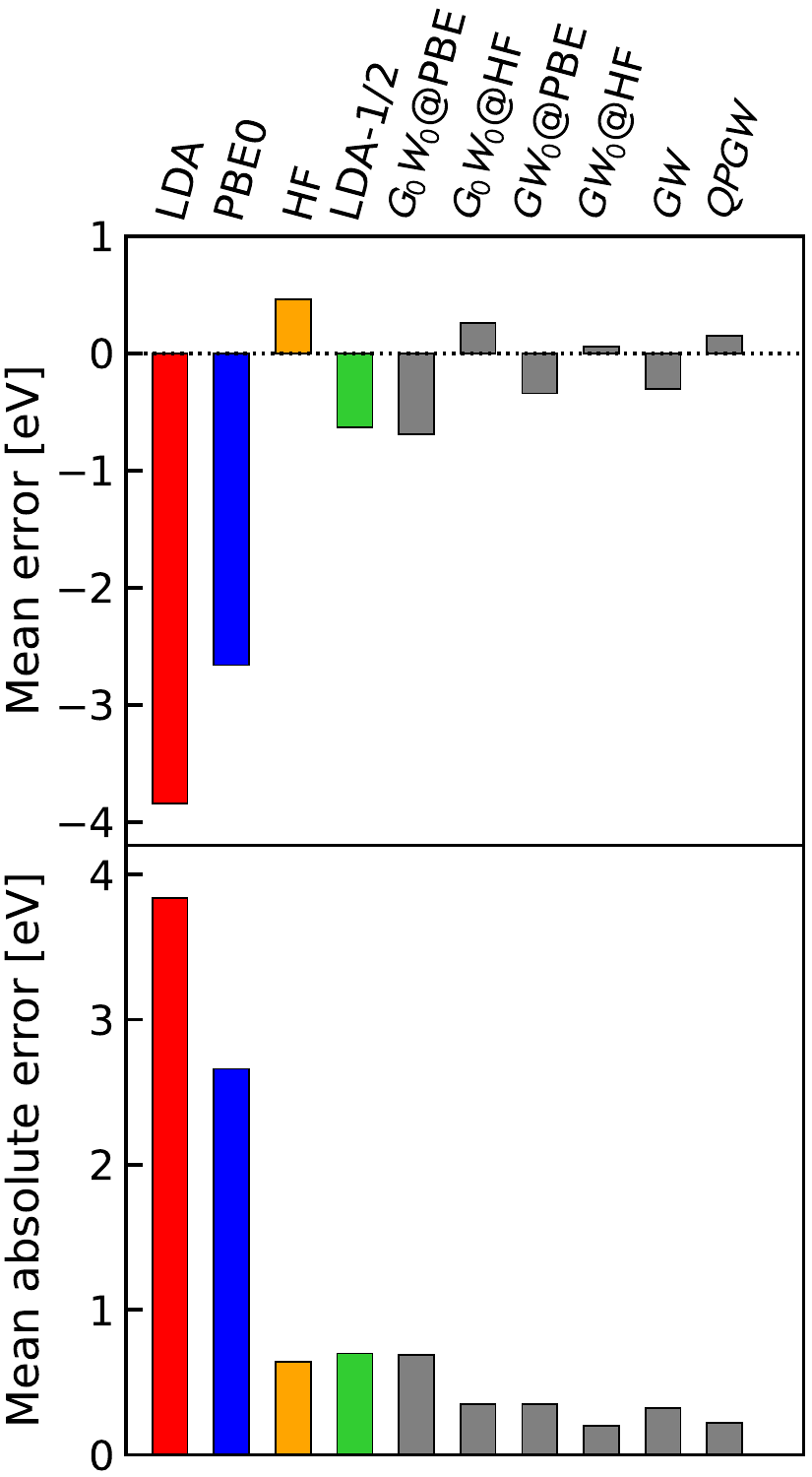}
\caption{Mean error (top) and mean absolute error (bottom) of HOMO energies of the of $GW100$ molecules obtained with a range of methods. The colored bars correspond to the data calculated in this paper, and the gray ones to the data from Ref. \citenum{Caruso2016}.}\label{fig-MEGW}
\end{figure}

Since both LDA-1/2 and HF yield HOMO energies with an accuracy similar to $GW$, it is interesting to compare our results with other $GW$ flavors reported by Caruso and coworkers.\cite{Caruso2016} This is done in Fig. \ref{fig-MEGW}, where we recognize that the methods can be grouped in three classes, according to their performance. The first class, with least accuracy, contains LDA and PBE0, which underestimate the IEs with a MAE larger than 2.0 eV. The second class, comprising HF, LDA-1/2, and $G_0W_0$@PBE, exhibits absolute MEs and MAEs between 0.4 and 0.7 eV. The third class includes the most accurate methods, $G_0W_0$@HF, $GW_0$@PBE, $GW_0$@HF, self-consistent $GW$, and quasi-particle self-consistent $GW$ ($QPGW$) with absolute MEs and MAEs smaller than 0.4 eV. It must be underlined that LDA-1/2 is the only method entirely based on a KS scheme with a local potential, while all other approaches with reasonable or good performance go beyond KS and employ non-local operators. 

\subsection{Planar vs. non-planar molecules}
To understand better how systematic the performance of LDA-1/2 is, we list the molecules whose IE is predicted with an absolute error larger than 1~eV: \ce{H2}, \ce{CI4}, \ce{Si2H6}, \ce{CBr4}, \ce{C4H5N3O}, \ce{F2}, \ce{AlI3}, \ce{GeH4}, \ce{C4H4N2O2}, \ce{CCl4}, \ce{Si5H12}, \ce{MgCl2}, \ce{CH4}, \ce{BH3}, \ce{SiH4}, \ce{C4H10}, \ce{MgF2}, \ce{C3H8}, \ce{B2H6}, \ce{TiF4}, \ce{SO2}, \ce{C2H6}, \ce{AlF3}, \ce{CF4}. On the other hand for \ce{He}, \ce{Na2}, \ce{H2O}, \ce{HN3}, \ce{NaCl}, \ce{K2}, \ce{GaCl}, \ce{BrK}, \ce{Rb2}, \ce{LiH}, \ce{CS2}, \ce{SH2}, \ce{HCl}, \ce{Li2}, \ce{LiF}, \ce{C2H4}, \ce{P2}, \ce{C6H6}, \ce{NH3} an absolute error smaller than 0.3~eV is obtained. Interestingly, all the systems belonging to the last group are linear or planar, except for \ce{NH3}. In contrast, in the former group, the majority of molecules is non-planar. This analysis suggests that the accuracy of LDA-1/2 can be related to the planarity of a molecule. We will explore this in the following.

\begin{figure}[htb]
\includegraphics{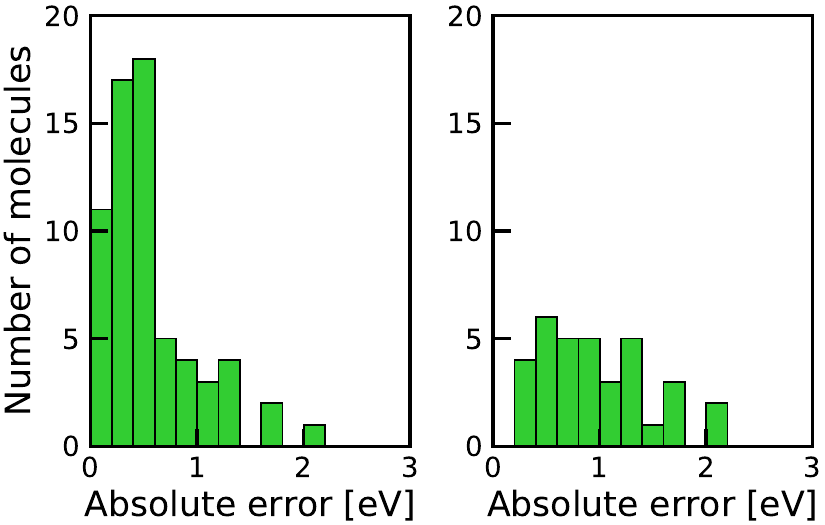}
\caption{Distribution of errors of HOMO energies in LDA-1/2 calculations for planar (left) and non-planar (right) molecules.}\label{fig-planar}
\end{figure}

In Fig. \ref{fig-planar}, we show the distribution of errors of LDA-1/2 for the HOMO energies of planar and non-planar molecules separately. As anticipated above, we observe that the non-planar molecules have a larger mean absolute error (0.97~eV) than the planar ones (0.56~eV). The MAE for these two subsets is displayed in Table \ref{tab-planar}, where it is compared to the MAE for all the molecules of the $GW100$ set. HF and LDA-1/2 share the same qualitative trend while the opposite trend can be recognized for LDA and PBE0. However, LDA-1/2 is the most sensitive method to the geometry of the molecule. This sensitivity is related to a similar one present in the $\Delta$SCF approach, one of the assumptions of the LDA-1/2 method. A previous investigation \cite{Shapley2001} based on $\Delta $SCF calculations with the LDA, demonstrated deviations from experiment for non-planar molecules to be larger than those for planar ones.

\begin{table}[htbp]
\caption{MAE (eV) of HOMO eigenvalues when all molecules of the $GW100$ set or subsets of only planar or non-planar molecules are taken into account.}\label{tab-planar}
\begin{tabular}{c|ccc}
\hline
Method	& All	& Planar	& Non-planar\\ \hline
LDA	& 3.84	& 3.93 &	3.68 \\
PBE0& 2.66	& 2.74&	2.50 \\
HF	& 0.64	& 0.60	& 0.72 \\
LDA-1/2	& 0.70	& 0.56&	0.97 \\ \hline
\end{tabular}
\end{table}

\subsection{Individual assessment of assumptions behind LDA-1/2}
\begin{figure}[htb]
\includegraphics{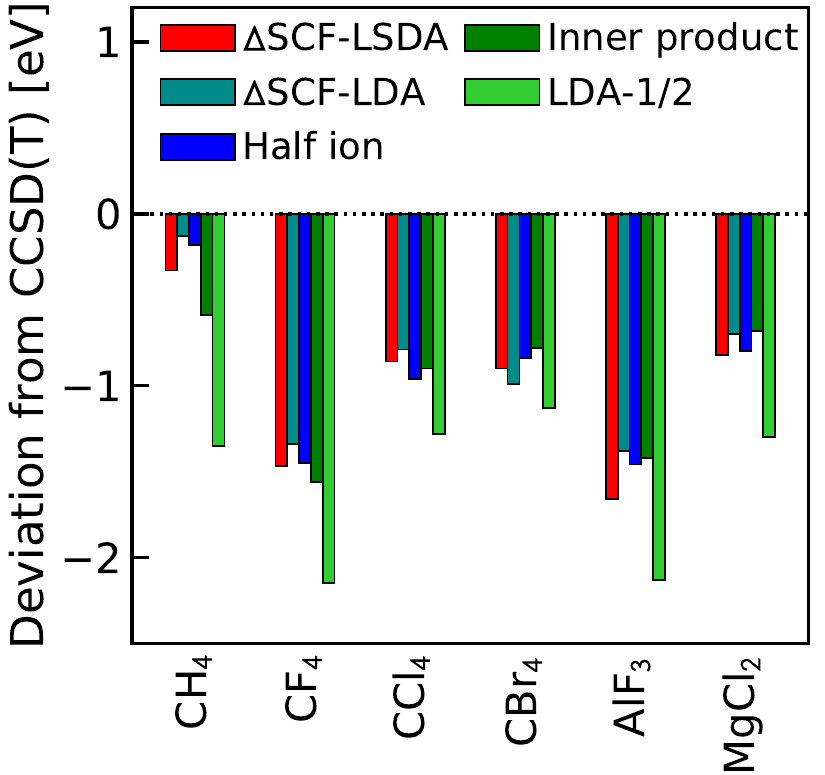}
\caption{Individual errors of the different assumptions behind the LDA-1/2 method.}\label{fig-deeper}
\end{figure}

We now turn our attention to the impact of the assumptions, listed in Section \ref{sec-lda05}, on the accuracy of the method. This analysis is summarized in Fig. \ref{fig-deeper} for the molecules \ce{CH4}, \ce{CF4}, \ce{CCl4}, \ce{CBr4}, \ce{AlF3} and \ce{MgCl2}, for which the deviation from the corresponding CCSD(T) result is larger than 1.1~eV. These significant errors should allow for identifying the origin of systematic errors. 

We first perform $\Delta$SCF calculations to obtain the IE. 
Since the removal of one electron leaves the molecule charged, and more difficult to converge with respect to the box size, $L$, the IE energy is finally obtained by extrapolation with the following expression:\cite{Makov1995}
\begin{equation}\label{eq.extr}
IE(L) = IE(\infty) + \frac{A}{L}.
\end{equation}
This procedure adds an uncertainty of about 0.1~eV.
%{\red until here}
%

Since an ionized molecule has an odd number of electrons, spin-polarized and non-spin-polarized $\Delta$SCF calculations yield different results, especially for molecules with a small number of electrons. The deviations between the approaches are shown in Fig. \ref{fig-deeper}. Since the differences are small, i.e., not larger than 0.3~eV, all the calculations discussed below are non spin-polarized.

The eigenvalues of the half-ionized molecules have also been obtained with an extrapolation procedure similar to Eq. (\ref{eq.extr}). The deviations with respect to CCSD(T) are given in Fig. \ref{fig-deeper}. We observe that the assumption of half-ionization does not introduce a large error over the IE obtained with the $\Delta$SCF approach, the largest difference between them being 0.17~eV. 

The deviations of the IEs obtained with the inner product given by Eq. (\ref{eq-inner}) are depicted in dark green. We note that what we call {\it inner product} contains actually a sum of two approximations: the one introduced by the inner product itself, and a second one, assuming $V_S$ as a superposition of atomic $V_S$. This is done mainly because it is difficult to disentangle them in practice. However, the approximation introduced by the inner product itself is expected to be smaller than the second one. We verify that, in all cases, except for \ce{CH4}, %{\red please, check the following} 
calculations under this assumption reproduce, with a small error, calculations of half-ionized molecules.

The last assumption included in LDA-1/2 is the neglect of the relaxation of KS orbitals when $V_S$ is introduced. We observe that this is the most drastic approximation among all the five, as it is leads to the most significant error.

%Finally, we depict in Fig. \ref{fig-deeper} the deviations of exact KS calculations, reported in Ref. \citenum{Chong2002}, with respect to CCSD(T). They are useful here to establish the limit of accuracy (with respect to CCSD(T)) which one could reach within the KS formalism.

\subsection{Low-lying KS levels}

\begin{figure*}[htb]
\includegraphics{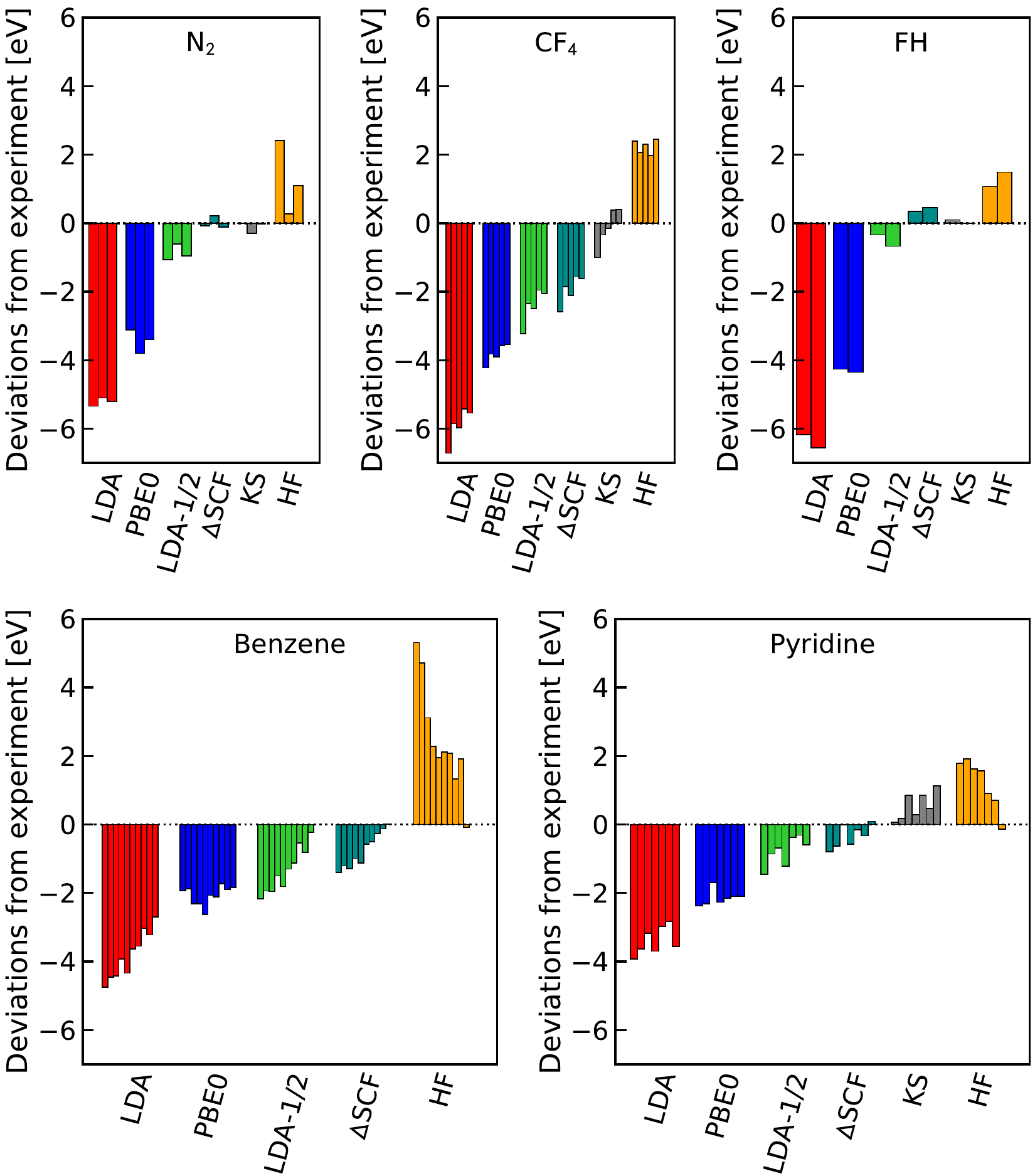}
\caption{Deviations of one-particle levels with respect to experiment. From the right to the left, HOMO, HOMO-1, HOMO-2, $\ldots$. KS and $\Delta$SCF results have been taken from Refs. \citenum{Chong2002} and \citenum{Shapley2001}, respectively.}\label{fig-morelevels}
\end{figure*}

Now we turn our attention to low-lying single-particle levels. In Fig. \ref{fig-morelevels}, we compare not only the HOMO but lower-lying single-particle levels with respect to experimental values, summarized in Ref. \citenum{Shapley2001}.\footnote{The ME and MAE between IEs obtained with CCDS(T) and the experimental IEs reported in Ref. \citenum{Shapley2001} are 0.004 and 0.08~eV, respectively. We proceed in this way, because in Ref. \citenum{Krause2015}, CCSD(T) calculations were carried out only for the first IEs.} Five molecules are chosen as a representative set that covers different types of structures and electrostatic properties. \ce{N2} is an example of a linear, non-polar molecule; \ce{CF4} is a representative of a non-planar molecule with polar bonds, but zero net dipole; \ce{FH} is a planar and highly polar molecule; benzene and pyridine represent aromatic, organic molecules; both of them are planar, but the first one has a null net dipole moment, whereas the second one has a non-vanishing net dipole. In Fig. \ref{fig-morelevels}, we also include results of exact KS and $\Delta$SCF calculations extracted from Refs. \citenum{Chong2002} and \citenum{Shapley2001}

We observe that exact KS calculations present the best agreement with experiment, followed by $\Delta$SCF calculations. In all cases except for pyridine, KS calculations tend to predict the HOMO energy with higher accuracy than low-lying levels (a very good discussion in this regard is provided by Ref. \citenum{Chong2002}). This same trend is also followed by $\Delta$SCF calculations, as already pointed out in Ref. \citenum{Shapley2001}. 

The accuracy of LDA-1/2 decreases as one goes down in energy. This is not surprising, since intrinsic limitations are inherited from one of its assumptions, namely that $\Delta$SCF calculations based on LDA can be used to obtain the HOMO energy. We recall that the LDA-1/2 corrections to the KS eigenvalues are meant specifically to improve the HOMO level. Strictly speaking, one should not expect that deeper levels are corrected appropriately as well. However, compared to LDA, we observe that even these deeper KS energies are improved. This can be attributed partially to the localization of charge obtained with the inclusion of the self-energy potential\cite{Pela2016}, one of the characteristics of LDA-1/2 which remedies the over-delocalization problem in LDA. On the other hand, $\Delta$SCF provides an upper bound for the accuracy of the LDA-1/2 method, and at the same time it is apparent from Fig. \ref{fig-morelevels} that differences between the LDA-1/2 and $\Delta$SCF results are small in comparison to the deviations from the experimental results. 

The performance of HF follows the same trends as $\Delta$SCF and LDA-1/2, i.e., the agreement with experiment deteriorates as one goes to deeper single-particle energies. This is observed in all cases, except for \ce{CF4}. In some cases, like in benzene and pyridine, this increase is much more pronounced, and, specifically for benzene, the deviation of HF eventually surpasses the absolute deviation of LDA, as one goes lower energies. Due to the opposite trends of HF and semi-local functionals (overestimation vs. underestimation), PBE0 has a discrepancy with respect to experiment which is roughly constant as one goes into low-lying energies.

\section{Conclusions}
We have presented the application of LDA-1/2 to atoms and the molecules of $GW100$ test set, evaluating the HOMO level with respect to CCSD(T) ionization energies. We have demonstrated that LDA-1/2 reaches a MAE similar to that of HF and $G_0W_0$@PBE, even although it employs only local potentials. The accuracy of the LDA-1/2 method is found to be better for planar molecules. Further, we have addressed the accuracy of the method in describing low-lying KS levels, and conclude that the HOMO eigenvalue is described most accurately. Since LDA-1/2 is a method employing exclusively local potentials, our findings may encourage other researchers to use this method for very large systems, where the computational cost may be prohibitive for methods based on non-local potentials.

\begin{acknowledgement}
R. R. Pela thanks the Alexander von Humboldt Foundation, Coordena\c{c}\~{a}o de Aperfei\c{c}oamento de Pessoal de N\'{i}vel Superior, and the Humboldt-Universit\"{a}t zu Berlin (grant ``Humboldt Talent Travel Award'') for financial support. 
This work was partially supported by the German Research Foundation (DFG) through the Collaborative Research Center 951 HIOS.
\end{acknowledgement}

\bibliography{references}
\end{document}